\documentclass[journal]{IEEEtran}
\usepackage{cite}

\ifCLASSINFOpdf 
\usepackage[pdftex]{graphicx}
\else
\usepackage[dvips]{graphicx}
\fi

\usepackage[cmex10]{amsmath}
\usepackage{algorithmic}
\usepackage[ruled,norelsize]{algorithm2e}

\usepackage{array}

\usepackage{mdwmath}
\usepackage{mdwtab}
\usepackage{todonotes}

\usepackage{eqparbox}

\ifCLASSOPTIONcompsoc
\usepackage[caption=false,font=normalsize,labelfont=sf,textfont=sf]{subfig}
\else
\usepackage[caption=false,font=footnotesize]{subfig}
\fi

\usepackage{fixltx2e}
\usepackage{url}
\usepackage{amsfonts}
\usepackage{amssymb}
\usepackage{amsthm}
\usepackage{bm}
\usepackage[ruled]{algorithm2e}
\usepackage[utf8]{inputenc}
\usepackage{booktabs}
\usepackage{url}
\usepackage{cite}
\usepackage{flushend}
\usepackage{epstopdf}

\newcounter{tempEquationCounter} 
\newcounter{thisEquationNumber}

\begin{document}
%
\title{Packet Duplication in Dual Connectivity Enabled 5G Wireless Networks: Overview and Challenges}


\author{Adnan~Aijaz,~\IEEEmembership{Senior~Member,~IEEE}
\vspace{-0.4cm}
\thanks{The author is with the Telecommunications Research Laboratory, Toshiba Research Europe Ltd., Bristol, BS1 4ND, UK. Contact e-mail: adnan.aijaz@toshiba-trel.com}}
\markboth{IEEE Communications Standards Magazine -- Accepted for Publication}%
{Shell \MakeLowercase{\textit{et al.}}: Bare Demo of IEEEtran.cls for Journals}
%


\maketitle
\begin{abstract}
\boldmath
Enabling ultra-reliable low latency communications (uRLLC) in 5G wireless networks creates challenging design requirements, particularly on the air-interface. The stringent latency and reliability targets require enhancements at different layers of the protocol stack. On the other hand, the \emph{parallel redundancy protocol} (PRP), wherein each data packet is duplicated and transmitted concurrently over two independent networks, provides a simple solution for improving reliability and reducing latency in wireless networks. PRP can be realized in cellular networks through the  \emph{dual connectivity} (DC) solution.  Recently, 3GPP has introduced packet duplication functionality in 5G wireless networks.  To this end, this paper provides an overview of the packet duplication functionality in 5G, in light of recent developments within 3GPP, and also highlights the related technical challenges.    

\end{abstract}



%
\IEEEpeerreviewmaketitle

\section{Introduction}
\IEEEPARstart{5}{G} wireless networks are expected to enable a range of use cases across different vertical industries. According to the International Telecommunication Union (ITU) \cite{ITU_2083}, such use cases can be broadly classified into three main service categories: enhanced mobile broadband (eMBB), massive machine type communications (mMTC), and ultra-reliable low latency communications (uRLLC). Based on these service categories, the Third Generation Partnership Project (3GPP) has recently identified the deployment scenarios and service requirements for next generation access technologies \cite{3gpp.38.913}. Among the three service categories, the most challenging design requirements arise for uRLLC, which encompasses critical applications like the \emph{Tactile Internet} \cite{TI_JSAC}. This is due to the stringent latency and reliability targets which become particularly difficult to achieve considering the  inherent trade-off between latency and reliability.  	 
Current 5G standardization activities reveal that the overall 5G radio access solution would consist of  \emph{evolved} long term evolution (LTE) radio access, complemented with backward-compatible enhancements, and \emph{New Radio} (NR) access technology operating in new spectrum bands. 

On the other hand, \emph{dual connectivity} (DC), introduced in Release 12 of 3GPP specifications \cite{3gpp.36.842}, allows a user to be simultaneously served by two different base stations, operating on different carrier frequencies, and connected via a non-ideal backhaul. Multi-connectivity extends the DC principle in the form of simultaneous connectivity to different radio access technologies (RATs).  DC primarily aims at improving per-user throughput and mobility robustness; however, it can potentially be exploited to improve the resilience of wireless transmissions. DC would be crucial in supporting tight interworking between LTE and NR access technologies. 

The design of air-interface is unarguably the most challenging aspect of enabling uRLLC over 5G wireless networks. In order to fulfil the stringent latency and reliability requirements, optimizations and enhancements are required at different layers of the air-interface protocol stack. To support uRLLC, a number of air-interface design characteristics, such as scaling of orthogonal frequency division multiplexing (OFDM) numerology, shortening of transmission time interval (TTI), optimization of user plane and control plane protocols, will come into play. The \emph{parallel redundancy protocol} (PRP), as specified in IEC 62439-3 \cite{IEC_PRP}, is particularly attractive for uRLLC. In PRP, every data packet is duplicated and transmitted concurrently over two independent networks. \textcolor{black}{Such packet duplication provides seamless redundancy that not only improves reliability but also reduces latency in communication by alleviating the need for retransmissions}.  Recently, 3GPP has introduced packet duplication functionality for 5G radio access as part of Release 15  \cite{3gpp.38.300}. \textcolor{black}{With DC and packet duplication, it becomes possible to realize PRP-like functionality in 5G wireless networks.}


\textcolor{black}{To this end, the main focus of this article is  packet duplication functionality in 5G wireless networks, in light of recent  3GPP Release 15 standardization activities. The key contributions are summarized as follows. }
\begin{itemize}
\item \textcolor{black}{We trace the evolution of packet duplication in wireless networks and provide a detailed description of packet duplication functionality in 5G.}

\item \textcolor{black}{We evaluate the performance of packet duplication in different scenarios through system-level simulations.}

\item \textcolor{black}{We highlight key technical challenges that arise by enabling packet duplication in 5G. }
\end{itemize}

\section{Evolution of Packet Duplication}
\begin{figure}
\centering
\includegraphics[scale=0.375]{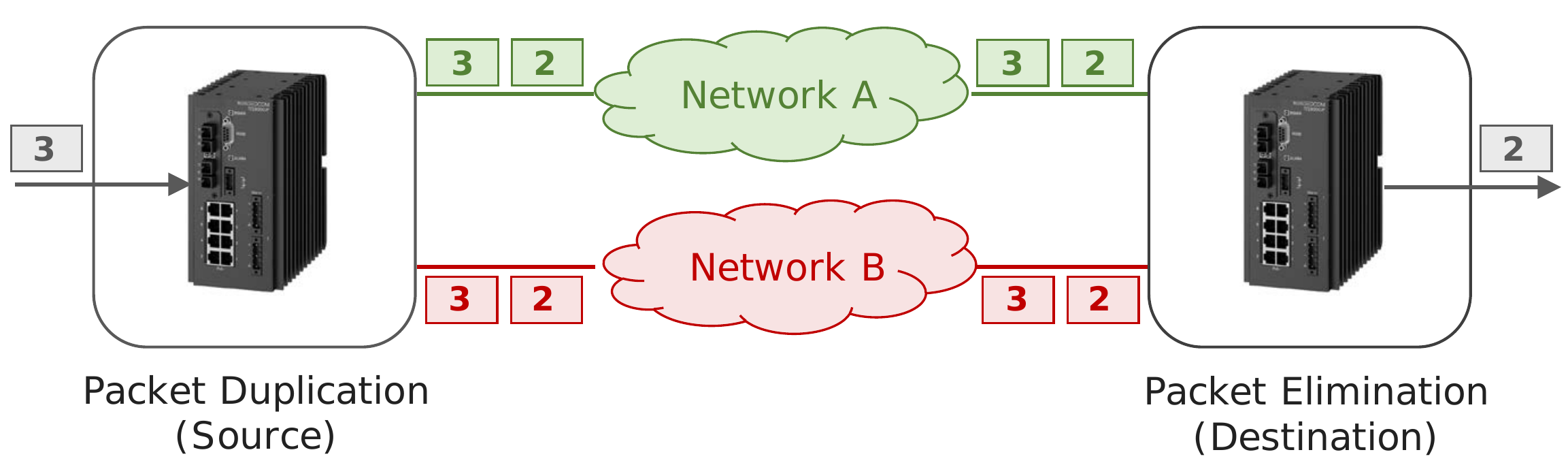}
\caption{Illustration of the PRP concept. Note that if a packet is lost over one network, it can still be recovered from the second network.}
\label{prp_func}
\end{figure}

The concept of packet duplication was first introduced in IEC 62439-3 specifications, in the form of PRP, to provide a certain degree of fault tolerance in industrial Ethernet networks. In case of PRP, every node is connected to two distinct networks. A source node employing PRP duplicates and transmits every data packet over two independent networks. The first copy that arrives at a destination node is retained whereas the second copy is discarded. PRP guarantees high availability for industrial Ethernet networks and provides seamless switchover and recovery in case of network failure. \textcolor{black}{This concept of PRP illustrated in Fig. \ref{prp_func} where a node is attached to two networks. A redundancy box takes care of duplicating packets at the transmitter side and removing duplicates at the receiver side. }

 PRP is particularly attractive for wireless environments as it provides a simple and robust solution to compensate for the effects of interference and small-scale disruptions due to which wireless networks are generally deemed unreliable. PRP provides seamless redundancy that improves reliability (and reduces latency)  as packets are lost (or retransmitted) only when they are dropped on both networks.  The concept of PRP can be applied to a range of wireless technologies.   Recently, Cena \emph{et al.} \cite{Wi-Red} proposed Wi-Red which is essentially seamless redundancy, as defined by PRP, applied to Wi-Fi networks. Similarly, Papadopoulos \emph{et al.} \cite{leapfrog} developed Leapfrog Collaboration which applies the PRP principle in the form of parallel transmissions over two paths in case of 6TiSCH wireless networks.  While PRP is under active investigation for different industrial wireless applications, its use in mobile/cellular networks is still at a nascent stage.  

\begin{figure*}
\label{comb_arch_stack}
\centering
\subfloat[]{\label{sys_arch}\includegraphics[scale=0.38]{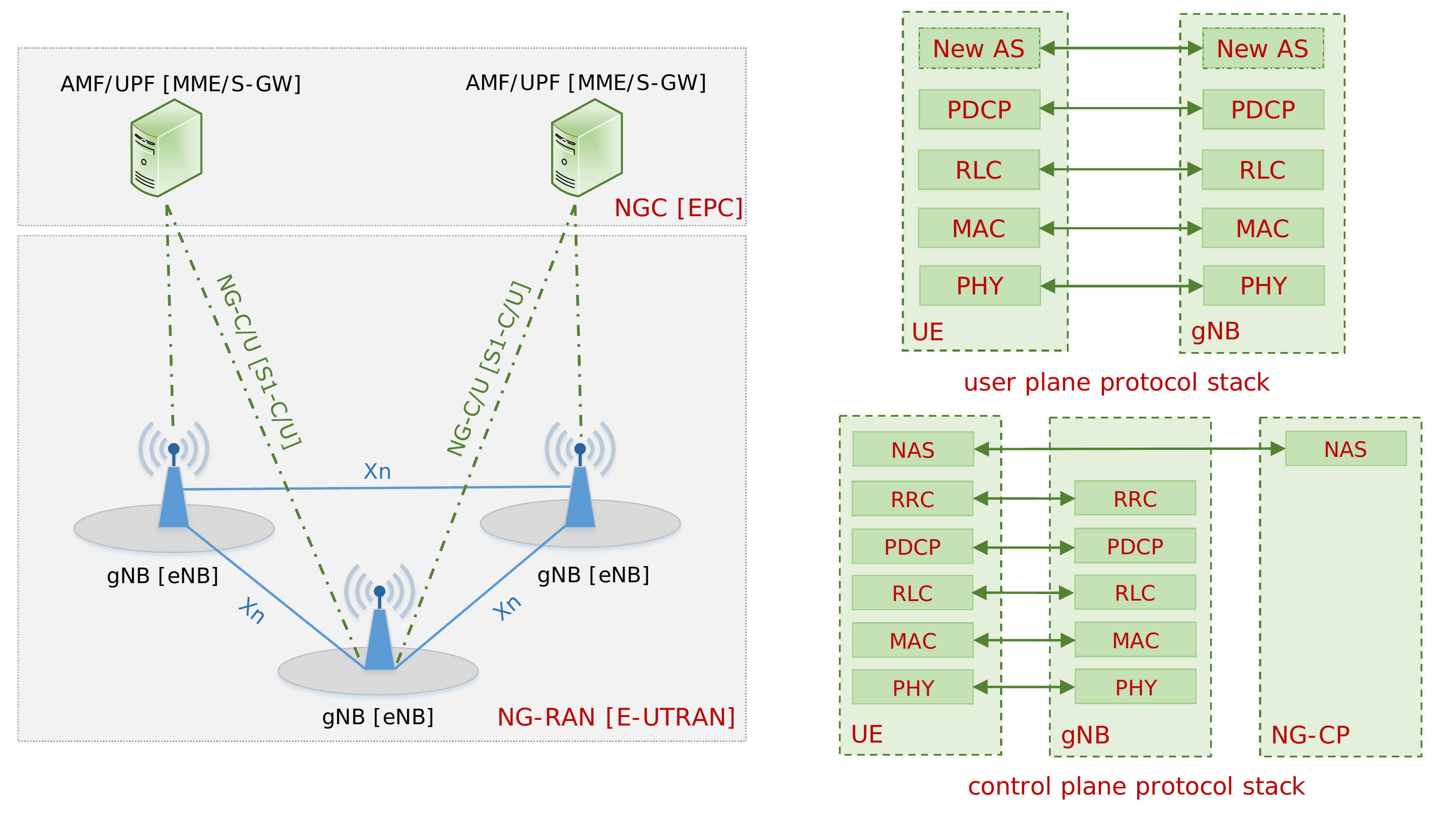}} \
\subfloat[]{\label{DC_stack}\includegraphics[scale=0.5]{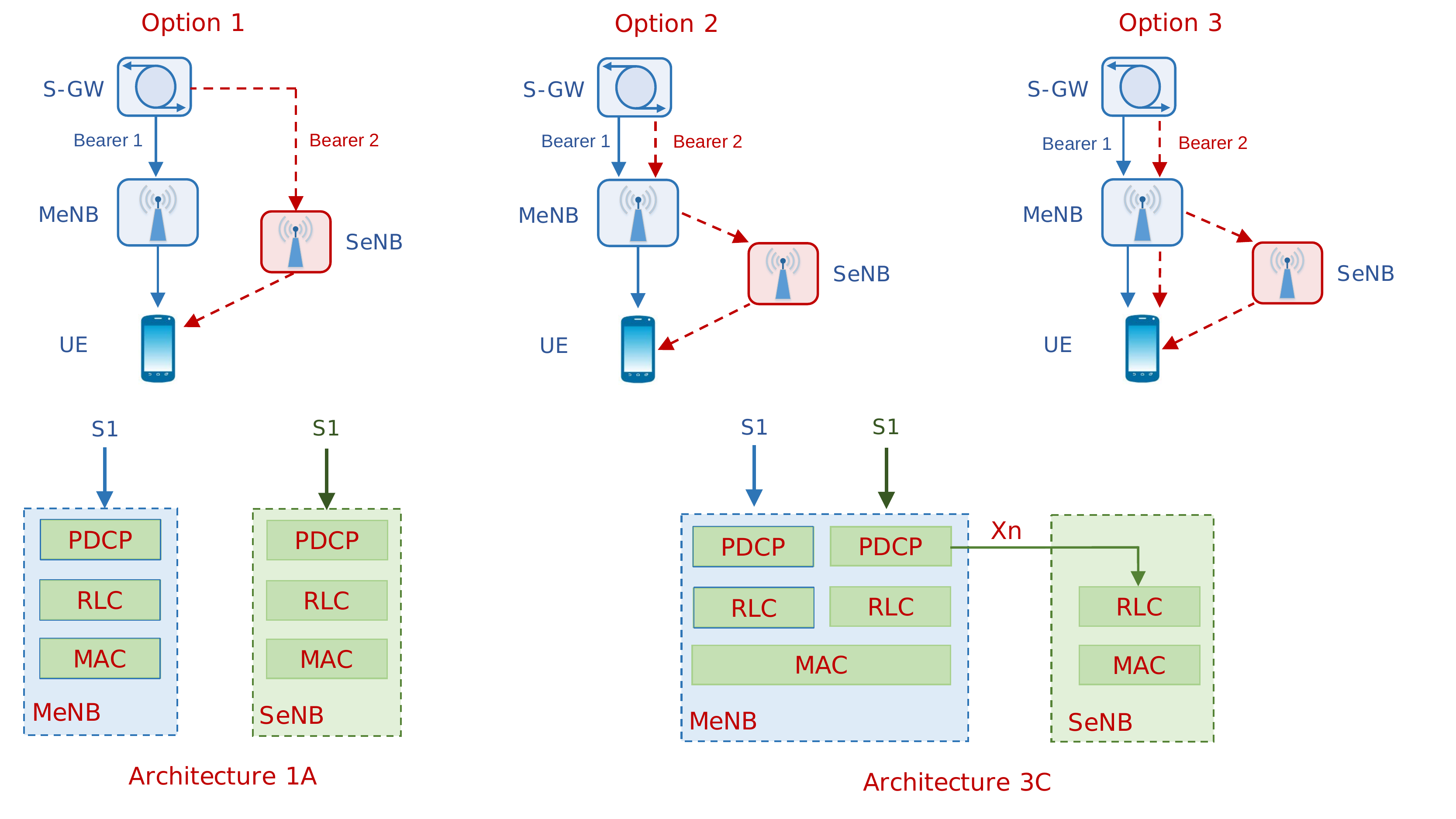}} \
\caption{(a) Overall system architecture for 5G-NR with respective LTE counterparts. The user plane and control plane protocol  stack for 5G-NR is also shown. (b) User plane data splitting options in DC and the protocol stack for different DC architectures.    }
\end{figure*}

\section{Background and Preliminaries} \label{prelim}
\subsection{Architecture and Protocol Stack}
The LTE radio access network (RAN), known as evolved universal terrestrial RAN (E-UTRAN), consists of the user equipment (UE), the evolved Node B (eNB), and the air-interface (E-UTRA). The eNBs are interconnected with each other by means of the X2 interface. The eNBs are connected by means of the S1 interface to the evolved packet core (EPC), more specifically to the serving gateway (S-GW) over the S1-U interface and to the mobility management entity over the S1-MME interface. 

The system architecture for next generation RAN (NG-RAN) \cite{3gpp.38.804} is shown in Fig. \ref{sys_arch}. The NG-RAN consists of the UE, the next generation Node B (gNB), and the air-interface which is termed as NG-RA. The gNBs are interconnected  by means of the Xn interface. The gNBs are also connected by means of the NG interface to the next generation core (NGC), more specifically to the access and mobility management function (AMF) over the N2 interface and to the user plane function (UPF) over the N3 interface. 

The user plane protocol stack for NR consists of the physical (PHY), medium access control (MAC), radio link control (RLC), and the packet data convergence protocol (PDCP) layers. In addition, a new access stratum (AS) layer has been introduced above PDCP. The key functionalities of different layers are described as follows. 
\begin{itemize}
\item \textbf{PHY} -- The PHY layer transmits all information from MAC transport channels over the air-interface and handles different functions such as power control, link adaptation and cell search. 

\item \textbf{MAC} -- The MAC layer provides mapping between logical channels and transport channels, and handles multiplexing/demultiplexing of RLC PDUs\footnote{PDU stands for protocol data unit and consists of upper layer service data unit (SDU) and the header. For example, MAC PDU = MAC SDU (RLC PDU) + MAC Header.}, scheduling information reporting, error correction, priority handling between UEs, transport format selection, etc. 

\item \textbf{RLC} -- The main functions of the RLC layer include transfer of upper layer PDUs according to transmission modes, error correction, \textcolor{black}{sequence numbering}, segmentation and re-segmentation, etc.  

\item \textbf{PDCP} -- The PDCP layer handles transfer of user data, header compression, sequence numbering, duplication detection, packet duplication, etc. 

\item \textbf{New AS} -- The new AS layer, which is termed as service data adaptation protocol (SDAP),  mainly handles mapping between a QoS flow and a data radio bearer.

\end{itemize}

In the control plane, PHY, MAC, and RLC layers perform the same functions as for the user plane. The PDCP layer performs ciphering and integrity protection. The control plane further consists of radio resource control (RRC) and non-access stratum (NAS) layers as described below.

\begin{itemize}
\item \textbf{RRC} -- The main functions of RRC layer include establishment, configuration, maintenance, and release of data radio and signaling radio bearers, addition, modification, and release of DC, broadcast of system information, mobility handling, etc.  

\item \textbf{NAS} -- The NAS layer mainly handles connection/session management functions between the UE and the NGC. 
\end{itemize}


\subsection{Dual Connectivity} \label{DC_options}
For the sake of describing the DC solution, we consider an LTE scenario and adopt the LTE terminology. With DC, a UE is simultaneously connected to two different base stations: a master eNB (MeNB) and a secondary eNB (SeNB). The MeNB and the SeNB are connected via a non-ideal backhaul and operate on different carrier frequencies. The group of serving cells associated with the MeNB and the SeNB is termed as master cell group (MCG) and secondary cell group (SCG), respectively.  DC is only applicable to UEs in RRC connected mode. A DC-enabled UE has two identities: one C-RNTI in the MCG and another C-RNTI in the SCG. \textcolor{black}{In case of DC, three different options can be distinguished for splitting of user plane data \cite{3gpp.36.842}.}

\begin{itemize}
\item \textbf{Option 1} -- User plane data is split in the core network, i.e., S1-U interface terminates in both the MeNB and the SeNB. In this case both eNBs have independent user plane connections towards the S-GW. 

\item \textbf{Option 2} -- User plane data is split in the RAN, i.e., S1-U interface terminates in the MeNB only \emph{without} bearer split in the RAN. In this case only the MeNB has a user plane connection towards the S-GW. 

\item \textbf{Option 3} -- User plane data is split in the RAN, i.e., S1-U interface terminates in the MeNB only \emph{with}  bearer split in the RAN. \textcolor{black}{Unlike Option 2, in this case packet-level split of user data is possible, i.e.,  data from one radio bearer can be transmitted from both eNBs. }

\end{itemize} 

These options are illustrated in Fig. \ref{DC_stack}. Based on these options, 3GPP has identified several user plane architectures for DC. \textcolor{black}{These architecture include 1A, 2A, 2B, 2C, 2D, 3A, 3B, 3C, and 3D \cite{3gpp.36.842}}. Here the numbers 1, 2, and 3 correspond to the three options discussed above and the alphabets A, B, C, and D correspond to independent PDCPs, master-slave PDCPs, independent RLCs, and master-slave RLCs, respectively. For realizing the DC solution, 3GPP has also standardized three different types of radio bearers: (i) MCG bearers (radio bearers served by the MeNB alone), (ii) SCG bearers (radio bearers served by the SeNB alone), and (iii) split bearers (radio bearers served by both the MeNB and the SeNB). In Release 12, 3GPP has agreed to support both 1A and 3C architectures for DC in the downlink and only 1A architecture in the uplink.  The support for bearer split in the uplink has been agreed in Release 13. From a control plane perspective, the MeNB is responsible for maintaining the RRC connection of a UE. The control plane connection towards the MME is always terminated in the MeNB. \textcolor{black}{For a detailed description of DC, interested readers are referred to \cite{DC_commag, 3gpp.36.842}.} Note that, DC can be realized in LTE-LTE, NR-NR, and LTE-NR scenarios.

\section{\textcolor{black}{Packet Duplication Functionality in 5G}}
\subsection{\textcolor{black}{PDCP Duplication}}
\textcolor{black}{3GPP RAN2 has recently introduced packet duplication functionality at the PDCP layer in 5G-NR \cite{3gpp.38.300}}. Packet duplication (PDCP duplication)  is supported for both user and control planes.  The PDCP layer in the transmitter is responsible for packet duplication whereas the PDCP layer in the receiver eliminates duplicate packets. The duplicated packet carries the same PDCP sequence number. The PDCP entity duplicates the PDU and not the SDU. The duplication of PDCP SDUs is not efficient as functions like header compression, ciphering, integrity protection, etc. are performed twice.  

\begin{figure*}
\centering
\includegraphics[scale=0.52]{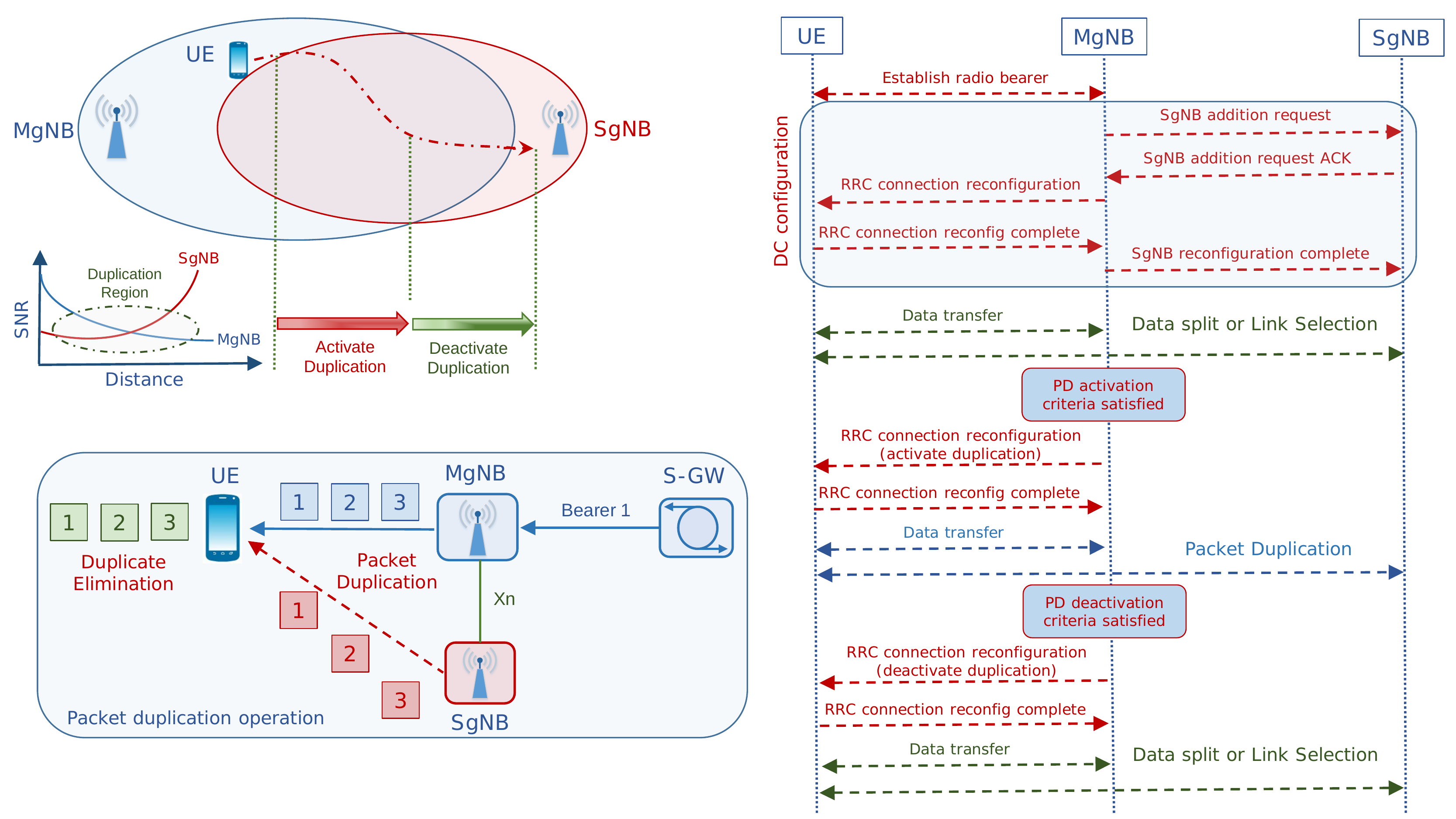}
\caption{Illustration of packet duplication operation in DC. Dynamic control of packet duplication through RRC signaling is also shown.  }
\label{pkt_dup}
\end{figure*}

Packet duplication in DC can be implemented with minimal impact via the split bearer architecture, which has been discussed earlier. In this case the duplication operation is similar to the split bearer operation. The only difference is that  the same PDCP PDU is transmitted via two separate RLC/MAC entities (also referred to as `legs'). Packet duplication can also be implemented by defining a new bearer type, e.g., duplicate bearer. The packet duplication operation is configured by the RRC layer. It can be configured at the UE level or at the radio bearer level. \textcolor{black}{However, the efficiency of packet duplication might change during the lifetime of a radio bearer \cite{TD_R21700172}}. Moreover, not all the applications require packet duplication. Therefore, it is recommended to configure the packet duplication operation at the radio bearer level.  When  duplication is configured for a radio bearer by RRC signaling, an additional RLC entity and an additional logical channel are added to the radio bearer to handle the duplicated PDCP PDUs. In case of DC, the two legs belong to different cell groups, i.e., MCG and SCG. When configuring the duplication operation, the RRC layer can also set the initial state of packet duplication, i.e., active or inactive.  Packet duplication may not always be beneficial during a bearer's lifetime. Its efficiency depends on a number of factors as explained later. Hence, dynamic control of packet duplication is desired, i.e, packet duplication must be dynamically activated or deactivated.  The dynamic activation/deactivation of packet duplication operation avoids unnecessary wastage of air-interface resources. 

\textcolor{black}{The packet duplication functionality can also be realized in the carrier aggregation (CA) scenario\footnote{Note that seamless redundancy through PRP is only possible  through packet duplication in DC scenario.} \cite{3gpp.38.300}}. Unlike DC, in CA user data is split in multiple carrier at the MAC layer. Similar to the DC case,  packet duplication is configured by the RRC layer. When duplication is configured for a radio bearer by RRC, an additional RLC entity and an additional logical channel are added to the original RLC entity and the logical channel pertaining to a radio bearer to handle the duplicated PDCP PDUs. However, there is a single MAC entity, as opposed to two separate MAC entities in case of DC.  It has been agreed in 3GPP RAN2 that PDCP duplication on the same carrier is not supported. Therefore, unlike the DC case, the mapping of the original and duplicate logical channels to different carriers also needs to be configured by the RRC layer. 3GPP RAN2 has agreed that packet duplication in CA is not supported if it is already configured in DC.

It is noteworthy that the PDCP layer in LTE already supports duplicate detection functionality based on the sequence number. Therefore, if the transmitter sends duplicate PDCP PDUs (via different legs), only the earlier received PDCP PDU can be processed at the receiver. The PDCP PDU arriving later is simply discarded without requiring any changes in the specification.  Hence, packet duplication can also be extended to the LTE-NR DC scenario. Proposals for introducing packet duplication functionality in LTE PDCP are under consideration within 3GPP.

\subsection{\textcolor{black}{Dynamic Control of PDCP Duplication}}

To achieve the dynamic control of packet duplication, different techniques can be employed. \textcolor{black}{One mechanism is to dynamically activate/deactivate packet duplication through RRC signaling \cite{3gpp.38.300}}. Based on a certain criteria (e.g., uplink/downlink channel conditions), the MgNB can activate/deactivate packet duplication through an RRC connection reconfiguration message. In some cases, the MgNB can provide the UE with criteria to activate/deactivate packet duplication. The UE evaluates the criteria to determine when to use packet duplication.    Despite its simplicity, this mechanism may incur significant overhead due to frequent signaling.  

Packet duplication can also be dynamically controlled through Layer 2 signaling. Since packet duplication is performed at the PDCP layer, a natural solution is dynamic activation/deactivation through a PDCP control PDU. This approach ensures the flexibility of activating/deactivating packet duplication at the radio bearer level. Another approach is to dynamically control packet duplication through a MAC control element (CE). The MAC CE approach allows activation/deactivation  commonly across all radio bearers configured with packet duplication by RRC. While the overhead of this approach is small, as compared to RRC-based control, it requires internal signaling between MAC and PDCP layers.


\textcolor{black}{The packet duplication operation in DC is illustrated through Fig. \ref{pkt_dup} which shows a UE moving from the center of MgNB to the center of SgNB. The UE is first configured with DC (i.e., addition of SgNB) through RRC signaling. As the user is moving towards the SgNB, the signal level from the MgNB reduces whereas that from the SgNB increases. Based on the trajectory of the UE, it is recommended to activate packet duplication during this period. Fig. \ref{pkt_dup} shows the activation of packet duplication through RRC signaling}. Once the criteria for packet duplication is satisfied (e.g., weak signal level from both the MgNB and the SgNB), the MgNB sends an RRC connection reconfiguration message to activate packet duplication. Once the UE sends the RRC connection reconfiguration complete message, the MgNB activates packet duplication. By using the split bearer architecture, same data is transmitted from both the MgNB and the SgNB. Note that upon activation of packet duplication, only the packets buffered at the PDCP layer are duplicated. The packets which have already been delivered to the RLC layer are not duplicated owing to significant complexity and the need for cross-layer interactions. As the UE moves towards the center of the SgNB, the channel conditions improve, and a single transmission might be sufficient. Hence, it is recommended to deactivate packet duplication. Once the criteria for deactivating packet duplication is fulfilled, the MgNB deactivates packet duplication through RRC signaling. Upon deactivation, the MgNB  might continue using the split bearer approach for data split. A link selection procedure can also be performed to determine the best cell for data transmission.

\subsection{\textcolor{black}{PDCP Duplication for Control Plane}}
\textcolor{black}{Packet duplication can also be applied for signaling radio bearers (SRBs)  to achieve robustness and RRC diversity \cite{TD_R21706630}}. Similar to the data bearers, packet duplication for SRBs is configured by the RRC layer. SRBs are characterized by small PDU sizes, less frequent transmissions, and higher scheduling priority. Moreover, different types of SRBs, which contain different control plane messages, have different priorities. Therefore, packet duplication for SRBs should be configured based on the type of the SRB (e.g., SRB1, SRB2). Further, in case of SRBs, the RRC command to activate/deactivate packet duplication can be embedded within an RRC control message. 

\begin{figure*}
\centering
\includegraphics[scale=0.5]{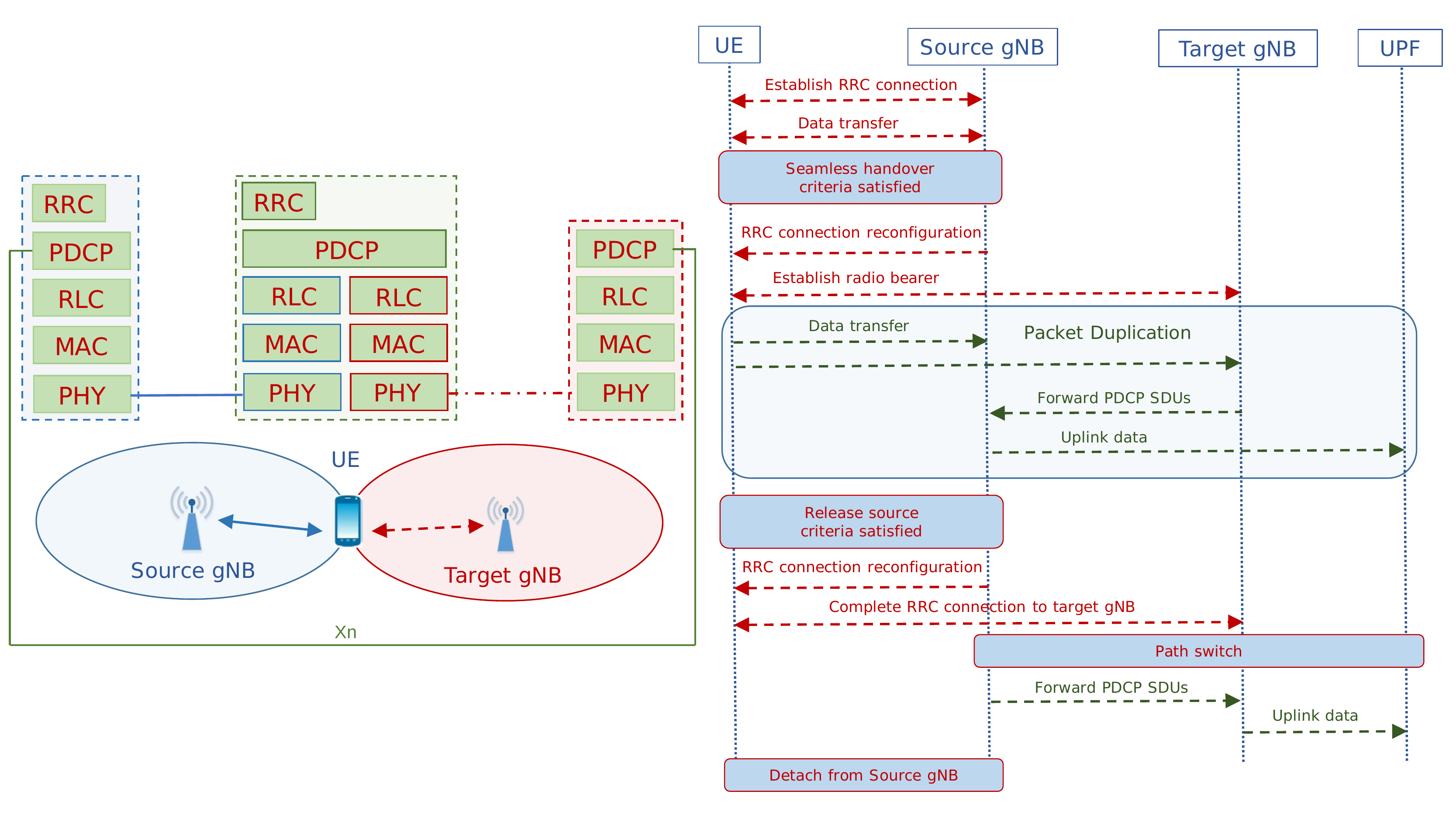}
\caption{Seamless handover procedure with packet duplication. The end-to-end signaling flow is also shown for the case of uplink packet duplication. The downlink scenario follows a similar procedure.    }
\label{ho_dup}
\end{figure*}

\subsection{\textcolor{black}{PDCP Duplication for Mobility Robustness}}
The packet duplication functionality is particularly attractive to improve the robustness of data and RRC signaling during handover procedure. Normally, a UE will only have one link for communication as it is required to release the RRC connection from the source gNB before it establishes a new RRC connection to the target gNB.  In some cases, a single link might not be sufficient to satisfy the target reliability. Simultaneous communication with both source and target gNBs  provides resilience to link failures. LTE already supports a make-before-break procedure which allows a UE to maintain connection with the source eNB even after receiving the handover command to establish a connection with the target eNB. Although this procedure reduces the mobility interruption time, it might not be sufficient to fulfil the requirements of uRLLC applications during mobility. 

\textcolor{black}{To improve mobility robustness for uRLLC applications, an enhanced make-before-break handover procedure has been proposed for 5G-NR \cite{TD_R21706710}. This handover procedure, which exploits PDCP duplication, is explained with the aid of Fig. \ref{ho_dup}. Note that in this case, there is a full protocol stack in both the source and the target gNBs. When the criteria for seamless handover is satisfied, the source gNB sends an RRC connection reconfiguration message to the target gNB to establish a radio bearer. Once a radio bearer to the target gNB is established, packet duplication can be used.  In the downlink case, the UE receives data from both the source and the target gNBs. The source gNB sends PDCP SDUs (along with the sequence numbers) to the target gNB over the Xn interface. Both gNBs separately perform header compression and encrypt the packets using corresponding keys. The received packets are dealt with individually at the UE. The PDCP entity in the UE is responsible for duplication elimination. In the uplink, the PDCP entity in the UE is responsible for duplicating PDCP SDUs. Each gNB individually treats the received PDUs. The target gNB is responsible for forwarding PDCP SDUs to the source gNB along with the sequence numbers. The source gNB is responsible for duplicate elimination until  the path switch is performed. The detailed signaling flow is also shown in Fig. \ref{ho_dup}.}

\subsection{\textcolor{black}{Efficiency of Packet Duplication}}
The efficiency of packet duplication is dependent on a number of factors which are described as follows. 
\begin{itemize}
\item \textbf{UE Mobility} -- The gain of packet duplication depends on the link quality which is affected by UE mobility.  Therefore, packet duplication is more effective if the UE is moving from the center of the gNB towards the cell-edge.   Packet duplication under good channel conditions might result in unnecessary wastage of  air-interface resources.  

\item \textbf{Latency of Xn Interface} -- DC is characterized by a non-ideal backhaul between the MeNB/MgNB and the SeNB/SgNB. Packet duplication may not provide benefit in terms of latency reduction if the latency of Xn interface is high. The packet via the SeNB/SgNB may arrive late at the receiver. In such cases, a retransmission from the MeNB/MgNB might be faster. 

\item \textbf{BLER Symmetry} -- The efficiency of packet duplication, in terms of reliability improvement, also depends on symmetry of block error rate (BLER) experienced by the MgNB/MeNB and SgNB/SeNB legs \cite{TD_R21700172}. Packet duplication is effective if both legs experience symmetric high BLER. In case of asymmetric BLER, a single transmission via the leg experiencing low BLER might be sufficient. 
\end{itemize}

\textcolor{black}{Therefore, packet duplication must be dynamically controlled and used only if the gain of duplication is expected. Otherwise, it would result in unnecessary overhead that may lead to degradation in network-level throughput. }

\section{Benefits of Packet Duplication}
Packet duplication  in DC-enabled 5G wireless networks provides a number of benefits which are described as follows.  

\begin{itemize}
\item \textcolor{black}{Packet duplication provides a simple solution toward meeting the stringent requirements of uRLLC}. It improves reliability by providing frequency and path diversity, and by compensating for individual packet losses due to radio link failures. It also  reduces latency by avoiding retransmissions at RLC and MAC layers or RRC connection re-establishment due to radio link failures.

\item Packet duplication potentially reduces jitter by minimizing the variance in latency. 


\item Packet duplication provides RRC diversity in the control plane which improves robustness for important signaling messages. 

\item Packet duplication improves handover performance through mobility robustness. 

\item Packet duplication is also beneficial for eMBB services. It can potentially improve throughput, particularly during TCP slow start phase. 

\item Packet duplication with link adaptation/selection improves radio resource utilization. 
\end{itemize}

\begin{table}
\caption{\textcolor{black}{Parameters for Performance Evaluation}}
\label{table1}
\vspace{-3ex}
\begin{center}
\begin{tabular}{ll}
\toprule
\bf{Parameter} & \bf{Value} \\
\midrule
Carrier frequency & \(5.2\) GHz\\
Downlink transmit power (Tier-1) & \(30\) dBm\\ 
Downlink transmit power (Tier-2) & \(23\) dBm\\ 
Uplink transmit power & \(18\) dBm\\ 
Standard deviation of shadowing  & $8$ dB (Tier-1), \(10\) dB (Tier-2) \\
Noise power & \(-174\) dBm/Hz\\
Noise figure & $5$ dB\\
Traffic model & Backlogged\\
Packet size & \(100\) bytes \\
\hline
\end{tabular}
\end{center}
\end{table}

\section{\textcolor{black}{Performance of Packet Duplication}}
\textcolor{black}{We evaluate the performance of packet duplication through system-level simulations in an indoor industrial environment. We consider a heterogeneous network deployment wherein two tiers of gNBs have been deployed. The first tier (Tier-1) consists of \(3\)-cell hexagonal grid model comprising microcells (or picocells).  The second tier (Tier-2) consists of picocells (or femtocells) which are randomly distributed within the coverage of first tier gNBs. The maximum cell radius for the first and second tiers is \(30\) meters and \(20\) meters, respectively.   We assume \(50\) uniformly distributed UEs in the coverage of each first tier gNB. \textcolor{black}{We assume that the UEs are static in nature and configured with PDCP duplication.}
For the sake of performance comparison, we define three distinct scenarios. The \emph{Scenario 1} refers to single-tier deployment with single-connectivity, i.e., user association with only Tier-1 gNBs. The \emph{Scenario 2} refers to two-tier deployment with single-connectivity such that user association is determined by maximum downlink received power. The \emph{Scenario 3} refers to two-tier deployment with DC. A UE is configured with DC only if it is within the coverage of both tiers of gNBs. We use the architecture 3C for DC in both downlink and uplink. We adopt an industrial propagation model \cite{ind_indoor}, based on which the path loss for distance (\(d\)) above \(15\) meters  is given by \(70.28+25.9 \times \log_{10} (d/15)\), and follows free-space model otherwise. The link-level model is based on standard signal-to-interference-plus-noise ratio (SINR). A transmission is successful if the received SINR is above a certain threshold \(\beta\). Other parameters are given in \tablename~\ref{table1}. We perform Monte Carlo simulations over different user and Tier-2 distributions with \(10^3\) packets (per user) in each iteration.   }

\begin{figure}
\label{comb_res}
\centering
\subfloat[]{\label{dl_dup}\includegraphics[scale=0.2]{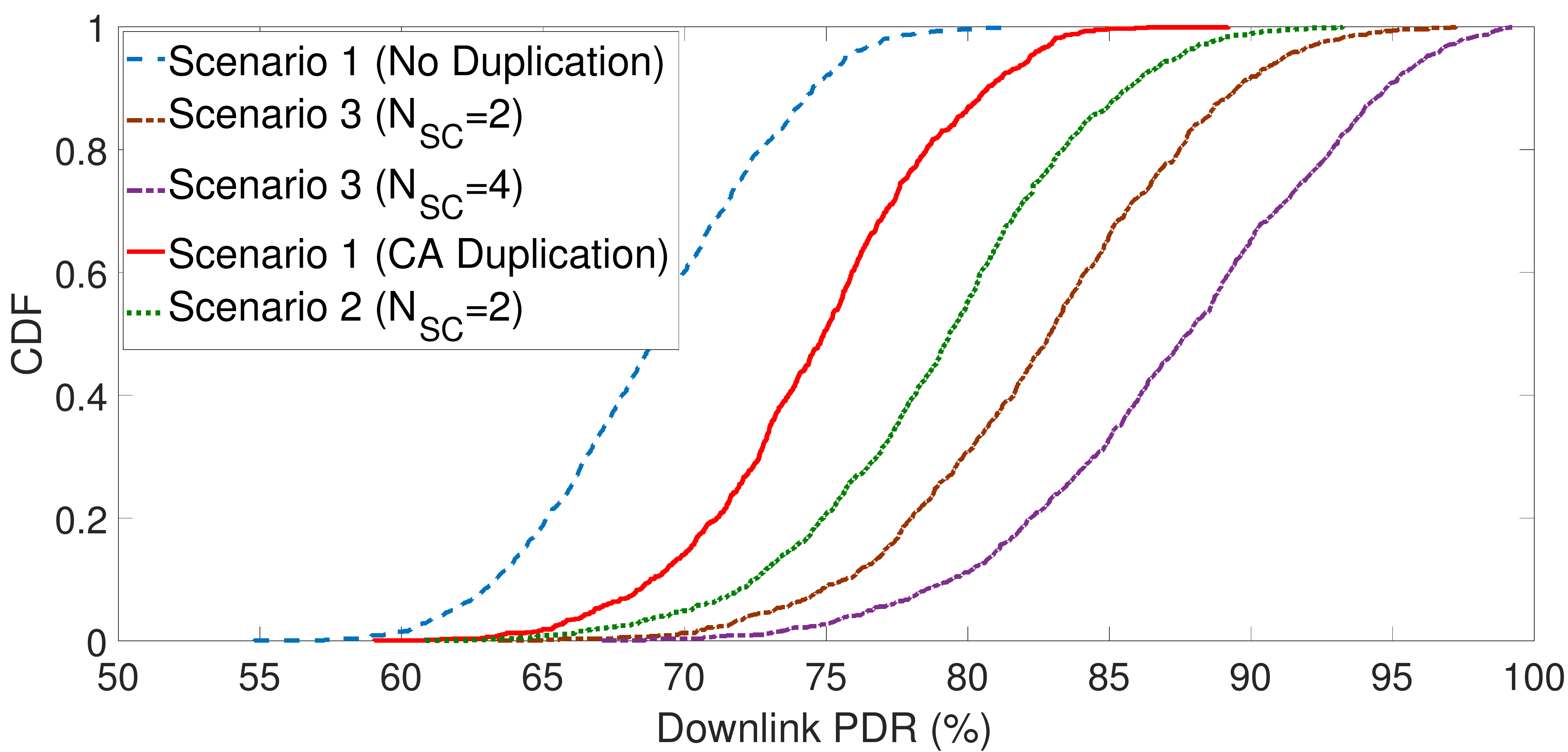}} \
\subfloat[]{\label{ul_dup}\includegraphics[scale=0.2]{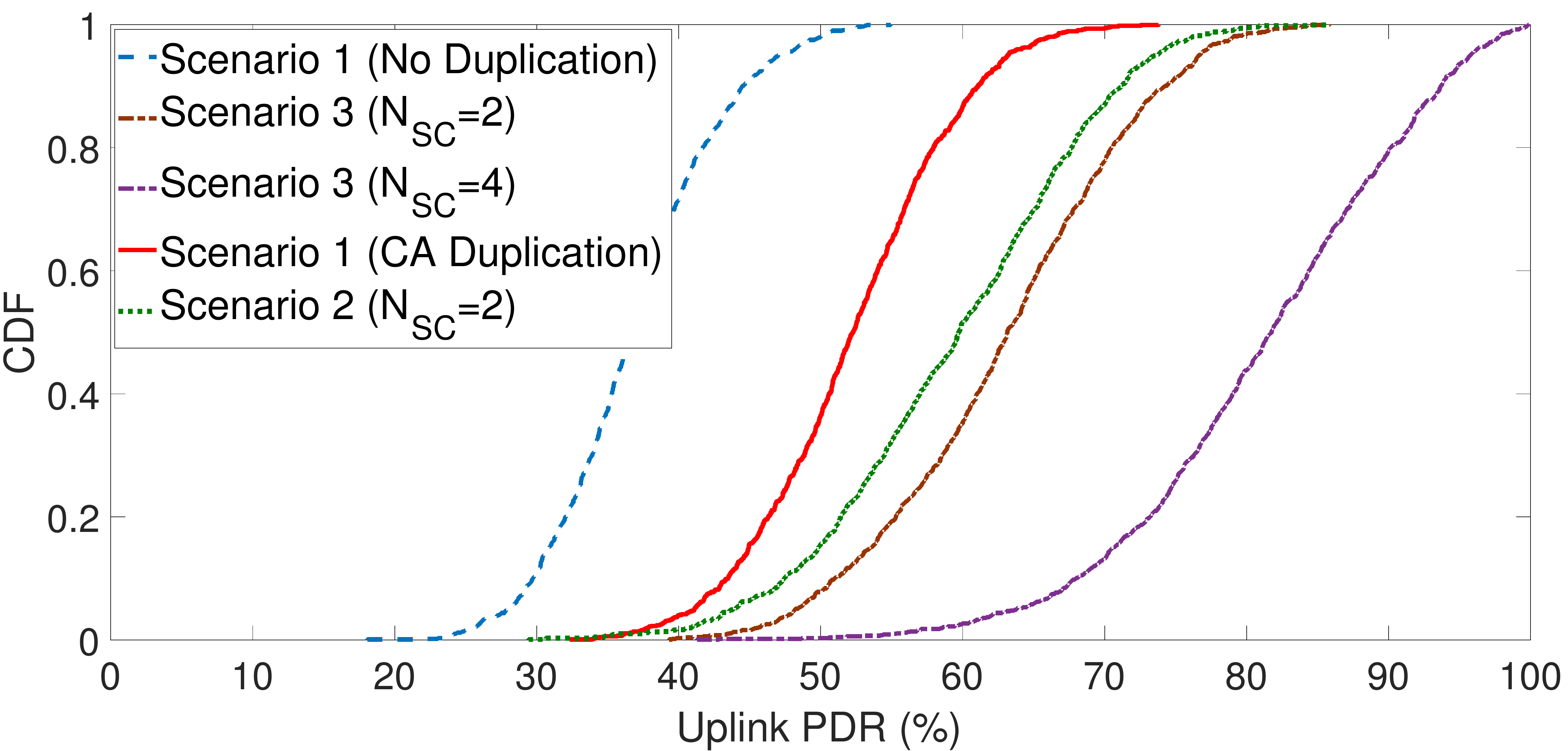}} \
\subfloat[]{\label{eff_dup}\includegraphics[scale=0.2]{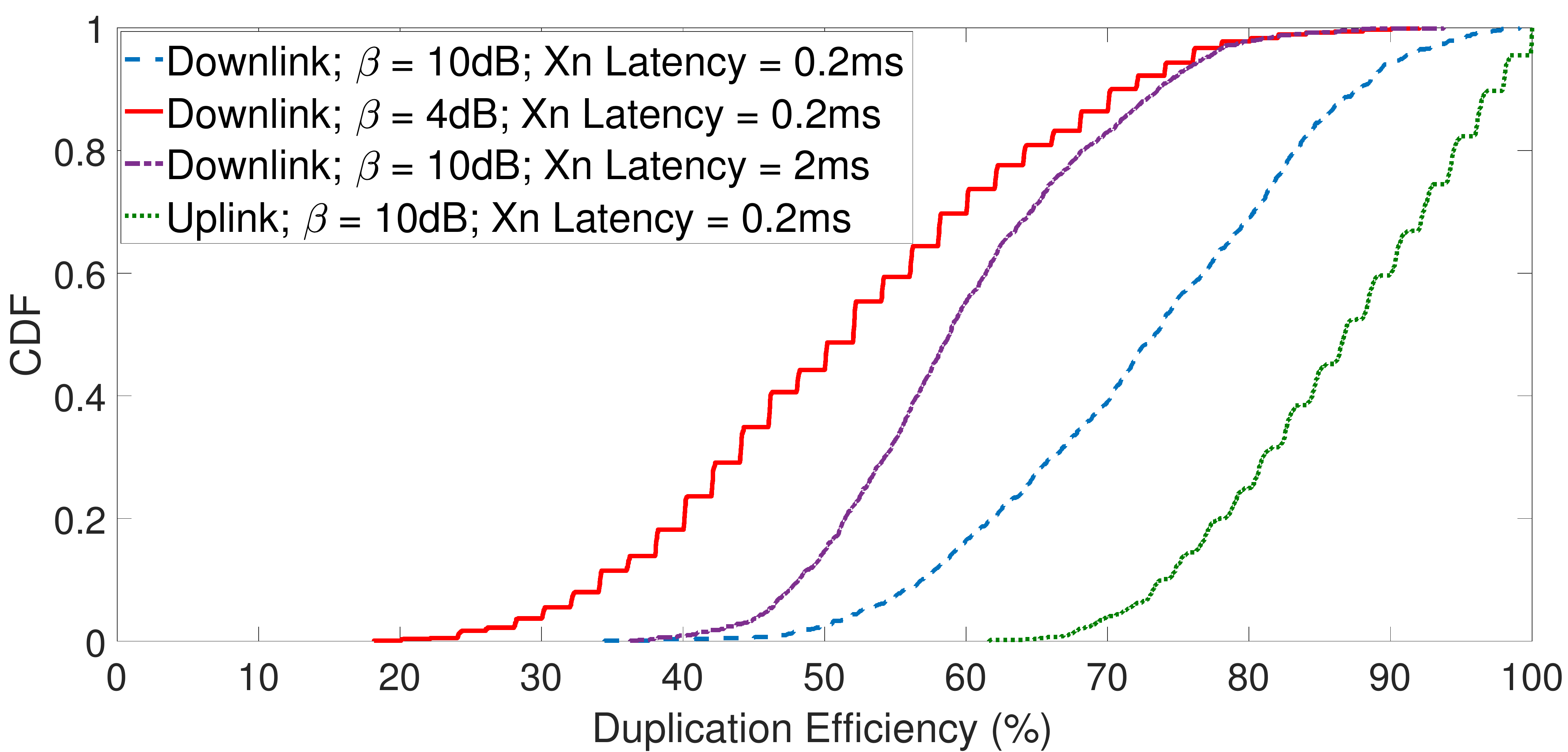}} \
\caption{\textcolor{black}{Performance evaluation of packet duplication: (a) CDF of downlink PDR; (b) CDF  of uplink PDR; (c) CDF of duplication efficiency (\(\text{N}_{\text{SC}}=2\)). CDF is generated over \(100\) iterations in all cases. \(\text{N}_{\text{SC}}\) denotes the number of Tier-2 gNBs per Tier-1 gNB. In both (a) and (b), \(\beta=10\) dB and one-way latency requirement is \(5\) ms.}   }
\end{figure}

\textcolor{black}{Fig. \ref{dl_dup} shows the cumulative distribution function (CDF) of downlink packet delivery ratio (PDR). The PDR is the ratio of successfully transmitted packets to the total number of packets. It can be easily inferred that packet duplication provides significant improvement in PDR performance. The baseline scenario of no duplication, i.e., \emph{Scenario 1} achieves a PDR of up to \(72\%\) in \(80\%\) of the cases. With packet duplication in DC, i.e., \emph{Scenario 3}, a gain of \(22\%\) is achieved in a similar setting with \(2\) second tier gNBs per first tier gNB. The achievable gain in PDR increases  with higher penetration of  second tier gNBs as it provide more opportunities for DC. Note that single-connectivity in two-tier deployment, i.e., \emph{Scenario 2} achieves higher PDR, and provide a gain of \(15\%\) as compared to the baseline scenario. This is mainly due to better channel conditions arising from improved coverage. Packet duplication in single-tier deployment, i.e, through duplication in CA, is also effective and outperforms the baseline scenario by providing a gain of \(8\%\). Fig. \ref{ul_dup} show the CDF of uplink PDR. The results follow a similar trend as the downlink case. However, the PDR performance is degraded as compared to the downlink case due to the power-limited nature of UEs in the uplink.     }

\textcolor{black}{We evaluate the effectiveness of packet duplication in terms of \emph{duplication efficiency} which refers to the percentage of transmissions for which duplication was inevitable, i.e., a single transmission was not sufficient. The duplication efficiency decreases under good channel conditions, i.e, when link-level SINR requirements are less stringent. For instance, duplication efficiency in downlink decreases by nearly \(20\%\) in \(80\%\) of the cases as \(\beta\) decreases from \(10\) dB to \(4\) dB. Moreover, packet duplication is  more effective in uplink than in downlink. The efficiency is also dependent on the latency of Xn interface. A higher latency reduces the effectiveness of packet duplication. The results demonstrate that dynamic control of packet duplication is particularly important to ensure efficient utilization of air-interface resources.  }

\section{Key  Challenges for Packet Duplication}
While packet duplication provides a number of benefits, it also creates various challenges which are described as follows. 

\subsection{Optimization of RLC Layer for PDCP Duplication}
When packet duplication is activated, the same PDCP PDU is transmitted via the MgNB leg (leg 1) and the SgNB leg (leg 2). Since the PDCP PDU in each leg is scheduled and transmitted independently, it might be successful in one leg and unsuccessful in the other leg. Consider that the transmission of the UE is successful via leg 1 whereas it failed in leg 2. We assume that the RLC entities in both the legs are operating in  acknowledged mode. Since the RLC entity of leg 2 is unaware of the successful transmission via the leg 1, it will request the UE to retransmit the lost RLC PDU, which results in an unnecessary retransmission in leg 1. This redundant retransmission results in wastage of air-interface resources. Besides, it blocks subsequent RLC PDU transmissions as retransmissions have higher priority over new PDUs. 

To resolve this issue, the RLC entity in the failed leg needs to be aware of the successful transmission in the other leg. This means cancelling a transmission on one leg when a duplicate packet is successfully transmitted  on the other leg. However, such interaction between the RLC entities requires intra-node and inter-node signaling. One solution \cite{TD_R21709498} is that the transmitting RLC entity in the successful leg informs the transmitting RLC entity in the failed leg, e.g., through an RLC sequence number. However, the RLC sequence number in both  legs needs to be aligned. If the transmitter is the gNB then some signaling mechanism is required over the Xn interface. In case the transmitter is a UE such information exchange becomes an implementation-specific issue. 


\subsection{PDCP Duplication versus Data Split in DC}
As mentioned earlier, packet duplication can be implemented in DC via the split bearer architecture. An important issue is whether a radio bearer can be simultaneously configured with PDCP duplication and data split. The data split approach is used to improve throughput. On the other hand, PDCP duplication is used to improve reliability and latency.  Moreover, data split and PDCP duplication cannot be simultaneously active. Apparently, there is no need to configure a radio bearer simultaneously with data split and PDCP duplication. Hence, DC may not be sufficient for those future applications requiring high throughput along with low latency and high reliability. 

\subsection{Duplicate Bearer versus Split Bearer}
The split bearer is configured with a split threshold and a path restriction which determines the path to be used when the PDCP data volume is less than the split threshold. The split threshold and the path restriction are ignored when packet duplication is activated; however, these must be taken into account upon deactivation. Such conditional application of the split threshold may lead to significant complexity from a UE implementation perspective \cite{TD_R21709095}. The complexity can potentially be reduced by defining a new type of radio bearer, e.g., duplicate bearer which is currently under discussion within 3GPP. The current proposal is to configure the duplicate bearer with packet duplication, the RLC entities pertaining to both legs, and the default leg. With duplicate bearer, both RLC entities are maintained once packet duplication is deactivated; however, the PDCP entity only transmits data to the default leg. 

\subsection{Packet Duplication with Implicit SCell Deactivation}
In case of CA, the gNB consist of multiple serving cells, one for each component carrier. The RRC connection is only handled by  one cell, called the primary cell (PCell) which is served by the primary component carrier. The other component carriers are referred to as the secondary cells (SCells). When CA is implemented in DC, both the MCG and the SCG consist of one or more  SCells.  A SCell can be implicitly deactivated (i.e., without informing the gNB) at the expiry of a timer called \textsf{sCellDeactivationTimer}. If one cell or all the cells pertaining to a leg are deactivated, the UE cannot transmit packets  on the deactivated cell or the deactivated leg, respectively. This leads to a conflict between packet duplication and implicit SCell deactivation. The gNB may activate packet duplication using the SCell which was implicitly deactivated or the SCell is implicitly deactivated while packet duplication is active. In both cases, the UE will not be able to transmit duplicated packets on the deactivated cell/leg unless it is activated. This may additionally lead to PDCP and/or RLC buffer overflow. 

\subsection{Packet Duplication with MgNB Handover}
With packet duplication in DC, the same PDCP PDU is transmitted via the MgNB and the SgNB. The mobility of UE has a direct implication on packet duplication. The packet duplication functionality is retained if a UE moves from a source SgNB to a target SgNB, as it is controlled by the MgNB.  However, packet duplication is dropped if a UE moves from a source MgNB to a target MgNB as the PDCP entity is located in the source MgNB. Hence, the legacy DC-based handover procedure needs to be enhanced to support packet duplication during MgNB handover event, if high reliability needs to be satisfied. The scenario becomes more complicated if the target MgNB decides to switch to a different SgNB.

\subsection{Enhancements for CA Duplication}
The packet duplication functionality in CA is realized through duplicating user data over multiple component carriers at the MAC layer. Unlike DC, there is a single MAC entity in CA. Although most of the existing CA procedures can be reused, some enhancements are required for realizing packet duplication. For instance, the association between each logical channel and the corresponding component carrier for original and duplication transmissions should be visible to the MAC layer. Moreover, when the UE performs logical channel prioritization, it needs to associate the logical channels to the corresponding uplink grants from different component carriers.



\section{Concluding Remarks}
3GPP has recently introduced packet duplication at the PDCP layer in 5G-NR. Packet duplication in DC-enabled 5G wireless networks provides seamless redundancy, which is crucial in satisfying the stringent user plane  requirements of uRLLC.  In the control plane, packet duplication improves robustness for signaling messages by providing RRC diversity. The dynamic control of packet duplication is particularly important in achieving the benefits of duplication without compromising air-interface resource efficiency. However, dynamic control must incur minimal signaling overhead.  This article described the packet duplication functionality in 5G-NR and highlighted the related technical challenges. Further evolution of packet duplication in 5G must address a range of issues pertaining to protocol optimizations with lower layer interactions and procedures.

\section{Acknowledgement}
The work presented in this paper is partly funded by the European Union’s Horizon 2020 research and innovation programme under grant agreement No 761745 and the Government of Taiwan. 
\bibliographystyle{IEEEtran}

\bibliography{IEEEabrv,pkt_bib}

\begin{thebibliography}{10}
\providecommand{\url}[1]{#1}
\csname url@samestyle\endcsname
\providecommand{\newblock}{\relax}
\providecommand{\bibinfo}[2]{#2}
\providecommand{\BIBentrySTDinterwordspacing}{\spaceskip=0pt\relax}
\providecommand{\BIBentryALTinterwordstretchfactor}{4}
\providecommand{\BIBentryALTinterwordspacing}{\spaceskip=\fontdimen2\font plus
\BIBentryALTinterwordstretchfactor\fontdimen3\font minus
  \fontdimen4\font\relax}
\providecommand{\BIBforeignlanguage}[2]{{%
\expandafter\ifx\csname l@#1\endcsname\relax
\typeout{** WARNING: IEEEtran.bst: No hyphenation pattern has been}%
\typeout{** loaded for the language `#1'. Using the pattern for}%
\typeout{** the default language instead.}%
\else
\language=\csname l@#1\endcsname
\fi
#2}}
\providecommand{\BIBdecl}{\relax}
\BIBdecl

\bibitem{ITU_2083}
\BIBentryALTinterwordspacing
ITU-R, ``{IMT Vision – Framework and Overall Objectives of the Future
  Development of IMT for 2020 and Beyond},'' {International Telecommunication
  Union (ITU)}, Recommendation ITU-R {M.2083-0}, Sept. 2015. [Online].
  Available:
  \url{https://www.itu.int/dms_pubrec/itu-r/rec/m/R-REC-M.2083-0-201509-I!!PDF-E.pdf}
\BIBentrySTDinterwordspacing

\bibitem{3gpp.38.913}
\BIBentryALTinterwordspacing
3GPP, ``{Study on Scenarios and Requirements for Next Generation Access
  Technologies},'' {3rd Generation Partnership Project (3GPP)}, TR {38.913},
  Oct. 2016. [Online]. Available:
  \url{http://www.3gpp.org/ftp/Specs/archive/38_series/38.913/}
\BIBentrySTDinterwordspacing

\bibitem{TI_JSAC}
M.~Simsek, A.~Aijaz, M.~Dohler, J.~Sachs, and G.~Fettweis, ``{5G-Enabled
  Tactile Internet},'' \emph{IEEE J. Sel. Areas Commun.}, vol.~34, no.~3, pp.
  460--473, March 2016.

\bibitem{3gpp.36.842}
\BIBentryALTinterwordspacing
3GPP, ``{Study on Small Cell Enhancements for E-UTRA and E-UTRAN: Higher Layer
  Aspects},'' {3rd Generation Partnership Project (3GPP)}, TR {36.842}, Dec.
  2013. [Online]. Available:
  \url{http://www.3gpp.org/ftp//Specs/archive/36_series/36.842/}
\BIBentrySTDinterwordspacing

\bibitem{IEC_PRP}
\BIBentryALTinterwordspacing
IEC, ``{Industrial Communications Networks -- High Availability Automation
  Networks -- Part 3: Parallel Redundancy Protocol (PRP) and High-Availability
  Seamless Redundancy (HSR)},'' {International Electrotechnical Commission
  (IEC)}, Standard (2nd Edition) {62439-3}, 2012. [Online]. Available:
  \url{https://webstore.iec.ch/publication/24447}
\BIBentrySTDinterwordspacing

\bibitem{3gpp.38.300}
\BIBentryALTinterwordspacing
3GPP, ``{NR and NG-RAN Overall Description},'' {3rd Generation Partnership
  Project (3GPP)}, TS {38.300}, Dec. 2017, {v2.0}. [Online]. Available:
  \url{http://www.3gpp.org/ftp/Specs/archive/38_series/38.300/}
\BIBentrySTDinterwordspacing

\bibitem{Wi-Red}
G.~Cena, S.~Scanzio, and A.~Valenzano, ``{Seamless Link-Level Redundancy to
  Improve Reliability of Industrial Wi-Fi Networks},'' \emph{IEEE Trans. Ind.
  Informat.}, vol.~12, no.~2, pp. 608--620, April 2016.

\bibitem{leapfrog}
G.~Papadopoulos \emph{et~al.}, ``{Leapfrog Collaboration: Toward Determinism
  and Predictability in Industrial-IoT Applications},'' in \emph{IEEE
  International Conference on Communications (ICC)}, May 2017.

\bibitem{3gpp.38.804}
\BIBentryALTinterwordspacing
3GPP, ``{Study on New Radio Access Technology; Radio Interface Protocol Aspects
  (Release 14)},'' {3rd Generation Partnership Project (3GPP)}, TR {38.804},
  March 2017. [Online]. Available:
  \url{http://www.3gpp.org/ftp/Specs/archive/38_series/38.804/}
\BIBentrySTDinterwordspacing

\bibitem{DC_commag}
C.~Rosa \emph{et~al.}, ``{Dual Connectivity for LTE Small Cell Evolution:
  Functionality and Performance Aspects},'' \emph{IEEE Commun. Mag.}, vol.~54,
  no.~6, pp. 137--143, June 2016.

\bibitem{TD_R21700172}
\BIBentryALTinterwordspacing
{Huawei}, ``{Evaluation on Packet Duplication in Multi-connectivity},'' {3rd
  Generation Partnership Project (3GPP)}, TDoc {R2-1700172}, Jan. 2017.
  [Online]. Available:
  \url{http://www.3gpp.org/ftp/tsg_ran/WG1_RL1/TSGR1_AH/NR_AH_1701/Docs/}
\BIBentrySTDinterwordspacing

\bibitem{TD_R21706630}
\BIBentryALTinterwordspacing
{Ericsson}, ``{Split SRB: Remaining Issue of Initial State, Path Selection and
  Duplication},'' {3rd Generation Partnership Project (3GPP)}, TDoc
  {R2-1706630}, Jun. 2017. [Online]. Available:
  \url{http://www.3gpp.org/ftp/TSG_RAN/WG1_RL1/TSGR1_AH/NR_AH_1706/Docs/}
\BIBentrySTDinterwordspacing

\bibitem{TD_R21706710}
\BIBentryALTinterwordspacing
{Huawei}, ``{Robust Data Transmission during Handover using Packet
  Duplication},'' {3rd Generation Partnership Project (3GPP)}, TDoc
  {R2-1706710}, Jun. 2017. [Online]. Available:
  \url{http://www.3gpp.org/ftp/TSG_RAN/WG1_RL1/TSGR1_AH/NR_AH_1706/Docs/}
\BIBentrySTDinterwordspacing

\bibitem{ind_indoor}
E.~Tanghe \emph{et~al.}, ``{The industrial indoor channel: Largescale and
  temporal fading at 900, 2400, and 5200 MHz},'' \emph{IEEE Trans. Wireless
  Commun.}, vol.~7, no.~7, pp. 2740 -- 2751, July 2008.

\bibitem{TD_R21709498}
\BIBentryALTinterwordspacing
Huawei, ``{RLC Optimization for Packet Duplication},'' {3rd Generation
  Partnership Project (3GPP)}, TDoc {R2-1709498}, Aug. 2017. [Online].
  Available: \url{http://www.3gpp.org/DynaReport/TDocExMtg--R2-99--17074.htm}
\BIBentrySTDinterwordspacing

\bibitem{TD_R21709095}
\BIBentryALTinterwordspacing
{LG Electronics Inc}, ``{Need for Duplicate RB},'' {3rd Generation Partnership
  Project (3GPP)}, TDoc {R2-1709095}, Aug. 2017. [Online]. Available:
  \url{http://www.3gpp.org/DynaReport/TDocExMtg--R2-99--17074.htm}
\BIBentrySTDinterwordspacing

\end{thebibliography}
%

\begin{IEEEbiographynophoto}{Adnan Aijaz}
(M'14--SM'18)   received the M.Sc degree in Mobile and Personal Communications and the Ph.D degree in Telecommunications Engineering from King’s College London (KCL), London, U.K., in 2011, and 2014, respectively. He is currently a Principal Research Engineer with the Telecommunications Research Laboratory of Toshiba Research Europe Ltd. His research broadly focuses on design and optimization of wireless networks with emphasis on cellular, Wi-Fi, cognitive, sensor, and mesh networks. 

Adnan has contributed to various national and international projects including UK Mobile VCE Core 5 (2011 – 2012), EU FP7 ACROPOLIS (2012 – 2013), EU FP7 SOLDER (2014 – 2015), UK Ofcom TV White Spaces Pilot (2014 – 2015), and EU H2020 Clear5G (2017 – 2020). He has been actively involved in various activities within IEEE. He was the symposium co-chair for IEEE SmartGridComm 2015. He was the co-organizer for the 5G and Tactile Internet (5G TACNET) workshop held at IEEE WCNC 2017 and IEEE ICC 2018. He is an associate editor for the IEEE Internet of Things Journal. He frequently serves as the TPC member on various flagship IEEE conferences. He is an active contributor to various standardization activities. Prior to joining KCL, he was with the cellular industry and worked in the areas of network performance management, optimization and quality assurance. 
\end{IEEEbiographynophoto}

\end{document}